\documentclass[11pt]{article}

\usepackage{arxiv}
\usepackage[utf8]{inputenc}
\usepackage[T1]{fontenc}
\usepackage{cite}
\usepackage{amsmath,amssymb,amsfonts}
\usepackage{algorithmic}
\usepackage[nolist]{acronym}
\usepackage{graphicx}
\usepackage{xcolor}
\usepackage{textcomp}
\usepackage{placeins}
\usepackage{xparse}
\usepackage[hidelinks]{hyperref}
\usepackage{url}
\usepackage{microtype}

\graphicspath{{gif/}}

\NewDocumentCommand{\figref}{ o m }{%
  Fig.~\ref{#2}%
  \IfValueT{#1}{(#1)}%
}

\newcommand{\secref}[1]{Sec.~\ref{#1}}
\newcommand{\valunit}[2]{\ensuremath{{#1\,\mathrm{#2}}}}
\newcommand{\singlefigwidth}{0.5\linewidth}
\newcommand{\ieeepreprintnotice}{%
  \begin{center}
    \footnotesize
    This work has been submitted to the IEEE for possible publication.
    Copyright may be transferred without notice, after which this version
    may no longer be accessible.
  \end{center}
}
\newcommand{\BibTeX}{{\rm B\kern-.05em{\sc i\kern-.025em b}\kern-.08em
    T\kern-.1667em\lower.7ex\hbox{E}\kern-.125emX}}

\title{A Fast and Energy-Efficient Latch-Based Memristive Analog Content-Addressable Memory}

\author{
Paul-Philipp Manea$^{1,2,3}$\\
\texttt{p.manea@fz-juelich.de}
\And
Aishwarya Natarajan$^{3}$
\And
Jim Ignowski$^{3}$
\And
John Paul Strachan$^{1,2}$
\And
Luca Buonanno$^{3}$
}

\preprintheader{Preprint submitted to IEEE JXCDC}
\date{May 12, 2026}

\begin{document}

\maketitle

\ieeepreprintnotice


\newacro{ADC}{analog-to-digital converter}
\newacroplural{ADC}[ADCs]{analog-to-digital converters}

\newacro{MLSA}{match line sense amplifier}

\newacro{DAC}{digital-to-analog converter}
\newacroplural{DAC}[DACs]{digital-to-analog converters}

\newacro{DRAM}{dynamic random-access memory}
\newacroplural{DRAM}[DRAMs]{dynamic random-access memories}

\newacro{SRAM}{static random-access memory}
\newacroplural{SRAM}[SRAMs]{static random-access memories}

\newacro{HBM}{high-bandwidth memory}
\newacroplural{HBM}[HBMs]{high-bandwidth memories}

\newacro{GPU}{graphics processing unit}
\newacroplural{GPU}[GPUs]{graphics processing units}

\newacro{CPU}{central processing unit}

\newacro{TPU}{tensor processing unit}
\newacroplural{TPU}[TPUs]{tensor processing units}

\newacro{IMC}{in-memory computing}
\newacroplural{IMC}[IMCs]{in-memory computing systems}

\newacro{CIM}{compute-in-memory}
\newacroplural{CIM}[CIMs]{compute-in-memory architectures}

\newacro{MAC}{multiply-accumulate}
\newacroplural{MAC}[MACs]{multiply-accumulate operations}

\newacro{VMM}{vector–matrix multiplication}
\newacroplural{VMM}[VMMs]{vector–matrix multiplications}

\newacro{KV}{key-value}
\newacroplural{KV}[KVs]{key-value pairs}

\newacro{LLM}{large language model}
\newacroplural{LLM}[LLMs]{large language models}

\newacro{RNN}{recurrent neural network}
\newacroplural{RNN}[RNNs]{recurrent neural networks}

\newacro{ANN}{artificial neural network}
\newacroplural{ANN}[ANNs]{artificial neural networks}

\newacro{AI}{artificial intelligence}

\newacro{GAI}{generative artificial intelligence}

\newacro{NLP}{natural language processing}

\newacro{PWM}{pulse-width modulation}
\newacroplural{PWM}[PWMs]{pulse-width modulations}

\newacro{QAT}{quantization-aware training}
\newacroplural{QAT}[QATs]{quantization-aware training strategies}

\newacro{ReLU}{rectified linear unit}
\newacroplural{ReLU}[ReLUs]{rectified linear units}

\newacro{CAM}{content-addressable memory}
\newacroplural{CAM}[CAMs]{content-addressable memories}

\newacro{aCAM}{analog content-addressable memory}
\newacroplural{aCAM}[aCAMs]{analog content-addressable memories}

\newacro{ML}{match line}
\newacroplural{ML}[MLs]{match lines}

\newacro{DL}{data line}
\newacroplural{DL}[DLs]{data lines}

\newacro{WL}{word line}
\newacroplural{WL}[WLs]{word lines}

\newacro{BL}{bit line}
\newacroplural{BL}[BLs]{bit lines}

\newacro{WE}{write enable}
\newacroplural{WE}[WEs]{write enable signals}

\newacro{MLSA}{match-line sense amplifier}

\newacro{PCSA}{precharge sense amplifier}

\newacro{PDK}{process design kit}

\newacro{SPICE}{simulation program with integrated circuit emphasis}

\newacro{CMOS}{complementary metal-oxide-semiconductor}

\newacro{MOSFET}{metal-oxide-semiconductor field-effect transistor}
\newacroplural{MOSFET}[MOSFETs]{metal-oxide-semiconductor field-effect transistors}

\newacro{NVM}{non-volatile memory}

\newacro{ReRAM}{resistive random-access memory}

\newacro{MRAM}{magnetoresistive random-access memory}

\newacro{PCM}{phase-change memory}

\newacro{FeRAM}{ferroelectric random-access memory}

\newacro{MTJ}{magnetic tunnel junction}

\newacro{FTJ}{ferroelectric tunnel junction}

\newacro{FeFET}{ferroelectric field-effect transistor}

\newacro{OSFET}{oxide semiconductor field-effect transistor}

\newacro{IGZO}{indium gallium zinc oxide}

\newacro{ITO}{indium tin oxide}

\newacro{HRS}{high-resistance state}

\newacro{LRS}{low-resistance state}

\newacro{SET}{set (programming voltage)}

\newacro{RESET}{reset (programming voltage)}

\newacro{FSM}{finite state machine}
\newacroplural{FSM}[FSMs]{finite state machines}

\newacro{LUT}{lookup table}
\newacroplural{LUT}[LUTs]{lookup tables}

\newacro{VTC}{voltage-to-time converter}
\newacroplural{VTC}[VTCs]{voltage-to-time converters}

\newacro{SDK}{software development kit}
\newacroplural{SDK}[SDKs]{software development kits}

\newacro{IP}{intellectual property}

\newacro{eFLASH}{embedded flash memory}

\newacro{3T1M}{three-transistor one-memristor}

\newacro{6T2M}{six-transistor two-memristor}

\newacro{8T2M}{eight-transistor two-memristor}

\newacro{SALM}{strong-arm latched memristor}

\newacro{LB}{low bound}

\newacro{HB}{high bound}

\newacro{GM}{memristor conductance}

\newacro{PE}{processing element}
\newacroplural{PE}[PEs]{processing elements}

\newacro{NOC}{network-on-chip}
\newacroplural{NOC}[NOCs]{networks-on-chip}

\newacro{RISC-V}{reduced instruction set computer version V}
\newacroplural{RISC-V}[RISC-Vs]{reduced instruction set computer version V processors}

\newacro{IRIS}{Iris flower dataset}

\newacro{SME}{small and medium enterprise}
\newacroplural{SME}[SMEs]{small and medium enterprises}

\newacro{BMFTR}{Bundesministerium für Forschung, Technologie und Raumfahrt}

\begin{abstract}
\Acp{aCAM} based on memristors provide a promising pathway toward energy-efficient large-scale associative computing for Edge AI and embedded intelligence applications. They have been successfully applied to decision-tree inference and extend the capabilities of \ac{CIM} architectures beyond conventional vector--matrix multiplication. However, conventional designs such as the 6T2M architecture suffer from static search power, limited voltage gain, and pronounced match-line crosstalk, constraining analog precision and scalability. We introduce a \ac{SALM} \ac{aCAM} cell that replaces static voltage division with a dynamic current-race comparator, enabling high regenerative gain, intrinsic result latching, and near-zero static search power. Compared to 6T2M, \ac{SALM} reduces read energy by 33\,\% at identical latency while eliminating the gain and crosstalk limitations that prevent 6T2M from scaling to large arrays. SALM further enables scalable \emph{sequential} and \emph{parallel} latch sharing, and a dataset-aware optimization framework exposes an explicit energy--latency tradeoff, achieving up to 50\,\% energy reduction at 3$\times$ latency across representative workloads. To enable architectural exploration, we develop a circuit-accurate behavioral model derived from SPICE lookup tables in 22\,nm FD-SOI technology, capturing match-line dynamics and crosstalk. Integrated into the X-TIME decision-tree compiler, this framework demonstrates that SALM maintains near-software accuracy for high-dimensional datasets, whereas baseline designs degrade due to limited gain and cumulative crosstalk.
\end{abstract}

\keywords{Memristors \and Analog Content-Addressable Memory \and Compute-in-Memory (CIM) \and Analog In-Memory Computing \and Edge AI \and Decision Trees \and Neuromorphic Hardware}

\begin{center}
$^{1}$PGI-14, Forschungszentrum Jülich, Jülich, Germany\\
$^{2}$RWTH Aachen University, Aachen, Germany\\
$^{3}$Hewlett Packard Enterprise, Fort Collins, CO, USA
\end{center}

\acresetall

\section{Introduction}

\subsection{Compute in Memory With Memristors}
\begin{figure}[htb!]
    \centering
    \includegraphics[width=\singlefigwidth]{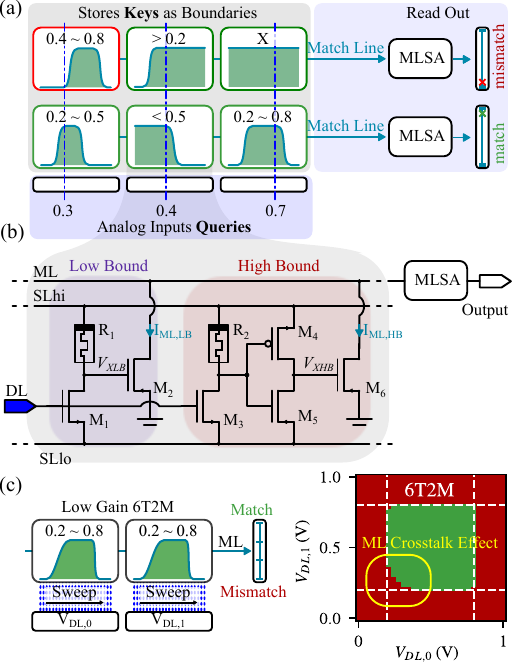}
    \caption{\textbf{(a)} Conceptual \ac{aCAM} operation. Each cell stores a lower and upper bound and compares an input query against this interval. A match is returned if the query lies within the boundaries indicated as the green area, otherwise a mismatch is produced. \textbf{(b)} Conventional \ac{6T2M} \ac{aCAM} cell implementing the window comparison with separate low-bound and high-bound branches that share a common match-line discharge path. \textbf{(c)} Experiment showing an input sweep applied to two aCAM cells with fixed stored boundaries. The resulting aCAM output highlights the impact of match-line (ML) crosstalk near the decision boundary.}
    \label{fig:into_fig}
\end{figure}

\Ac{CIM} has introduced a new architectural paradigm to address memory bottlenecks, particularly suited for \ac{AI} workloads. In one embodiment, highly parallel and energy-efficient crossbar arrays execute operand-heavy \acp{VMM} directly within the memory fabric, constituting the dominant computational workload in neural network architectures. As a result, only the computed outputs must be transferred out of the memory array rather than entire weight matrices, significantly reducing data movement, energy consumption, and latency \cite{sebastian2020memory, wan2022compute, leroux2025analog}.

Among emerging non-volatile memory devices, memristors—often realized as resistive random-access memory (RRAM)—have emerged as a leading candidate for implementing \ac{CIM} architectures due to their compact footprint, analog programmability, and compatibility with dense crossbar integration \cite{strukov2008missing}. These two-terminal devices store information by switching between distinct resistance states through the controlled formation and rupture of conductive filaments in a dielectric layer \cite{waser2009redox}. Their nanoscale cell size, high integration density, low-power operation, and multi-level conductance capability enable analog computation directly within memory arrays \cite{sebastian2020memory}.

\Acp{CAM} are a form of associative memory specifically designed for high-speed search and pattern-matching operations. Traditionally studied as a separate memory primitive, they can also be viewed as a complementary realization of \ac{CIM}, since comparisons are executed directly within the memory array rather than in a separate processing unit. They perform parallel comparisons between input queries and stored keys directly in hardware, returning the address of any matching key \cite{1599540}. Conventional \acp{CAM} operate on binary or ternary keys (TCAM) and were initially designed predominantly for high-speed lookup and forwarding of Internet Protocol (IP) packet headers in network routers \cite{pei1991vlsi}.

Unlike TCAMs, \acp{aCAM} store and compare analog values by exploiting the multi-level conductance tunability of memristors, allowing each cell to determine whether an input voltage lies within a programmable analog range \cite{Li2020}, which is graphically illustrated in \figref[a]{fig:into_fig}. This capability enables continuous-valued and multi-interval matching, making \acp{aCAM} well suited for similarity-based tasks within \ac{AI} and signal-processing \cite{10843122}. These include Decision Trees and Random Forests \cite{pedretti2021tree}, Finite State Machines \cite{Graves2022}, DNA pattern matching \cite{he2024shiftcam}, similarity calculations within transformer Attention Mechanisms \cite{maena_gain_cell_aCAM}, and even to replace power-hungry \acp{ADC} in \ac{CIM} systems \cite{Zhao2023RACEITAR}.

The earliest and most widely studied \ac{aCAM} architecture is the \ac{6T2M} design, shown in \figref[b]{fig:into_fig} \cite{Li2020}. It consists of two compartments for the \ac{LB} and \ac{HB} inequality checks, each implemented by a 1T1R voltage divider formed by one transistor and one memristor in series. For the \ac{LB} comparison, the divider node directly drives the \ac{ML} pull-down transistor ($\textrm{M}_\textrm{2}$), whereas for the \ac{HB} comparison it is routed through an inverter stage that flips the decision polarity and controls a second pull-down transistor ($\textrm{M}_\textrm{6}$). Upon a mismatch, either pull-down device ($\textrm{M}_\textrm{2}$ or $\textrm{M}_\textrm{6}$) discharges the \ac{ML}, whose voltage is evaluated by an external \ac{MLSA} to determine the row-level match condition. Thus, the \ac{ML} realizes a NOR-like aggregation across all cells in a row: if any inequality check in any cell mismatches, the \ac{ML} discharges and produces a global mismatch for the stored word. The key advantage of \ac{6T2M} is its low component count, which simplifies integration and reduces cell area.

The primary limitations of the \ac{6T2M} cell are \textit{static energy consumption}, \textit{low output gain}, and \textit{match-line crosstalk}. Together, these mechanisms constrain energy efficiency and array scalability. In particular, the limited match--mismatch separation prevents nominally matched cells from remaining fully passive, so the row decision can be influenced by cells that should not discharge the \ac{ML}. These mechanisms are discussed in more detail below.

\vspace{-0.2cm}
\paragraph{Static Energy Consumption}
The 1T1R voltage divider draws continuous static current during search operations. In addition, when the \ac{HB} branch inverter operates at intermediate voltages, both $\textrm{M}_\textrm{4}$ and $\textrm{M}_\textrm{5}$ conduct simultaneously, generating additional short-circuit current.

\vspace{-0.2cm}
\paragraph{Low Output Gain}
The \ac{6T2M} cell exhibits low effective voltage gain at the pull-down nodes $V_{XLB}$ and $V_{XHB}$. This shallow input--output characteristic, shown in \figref[c]{fig:fig1}, limits the number of reliably distinguishable analog levels~\cite{Bazzi2022} and weakens match--mismatch separation. As a result, the \ac{ML} voltage can settle near the \ac{MLSA} threshold, reducing the sensing margin.

\vspace{-0.2cm}
\paragraph{Match Line Crosstalk}
Due to limited match-mismatch separation, nominally matched cells can still contribute residual pull-down currents instead of remaining electrically passive. In large rows, these parasitic contributions accumulate and perturb the shared \ac{ML} voltage. As shown in \figref[c]{fig:into_fig}, the resulting output surface deviates from the ideal dashed boundary and exhibits a joint dependence on multiple inputs rather than depending only on the mismatching cell. This effect causes localized distortions near the decision boundary and constitutes a fundamental scalability bottleneck for conventional 6T2M \ac{aCAM} arrays.

Motivated by these limitations, the main contributions of this work are:

\begin{itemize}
    \item A Strong Arm Latched Memristor (SALM) aCAM cell that eliminates static search current while providing high regenerative gain and intrinsic result latching.

    \item A scalable latch-shared SALM aCAM architecture enabling configurable sequential and parallel scaling of analog inequalities, supported by an application-driven optimization framework for latency, area, and energy trade-offs.

    \item A circuit-accurate behavioral aCAM model derived from SPICE lookup tables that captures match-line discharge dynamics and crosstalk for large-scale inference evaluation.
\end{itemize}

The remainder of this paper is organized as follows. In \secref{sec:SALM}, we introduce the SALM aCAM cell and describe its dynamic comparison principle, followed by a circuit-level evaluation of gain, energy, latency, and area relative to the 6T2M design. In \secref{sec:SALM_architecture}, we analyze array-level scaling through sequential and parallel reuse of the latch and derive analytical models enabling architecture optimization for a given application and technology. To support system-level evaluation beyond transistor-level simulation, \secref{sec:beh_model} presents a circuit-accurate behavioral model that reproduces match-line discharge dynamics and crosstalk effects. Finally, in \secref{sec:re_inference}, we integrate this model into the X-TIME compiler flow and quantify inference accuracy across different aCAM architectures on representative Random Forest datasets. We further analyze how the architectural optimization framework introduced in \secref{sec:SALM_architecture} impacts energy consumption across datasets and characterize the resulting latency--energy trade-offs.

\begin{figure*}[htb!]
    \centering
    \includegraphics[width=\textwidth]{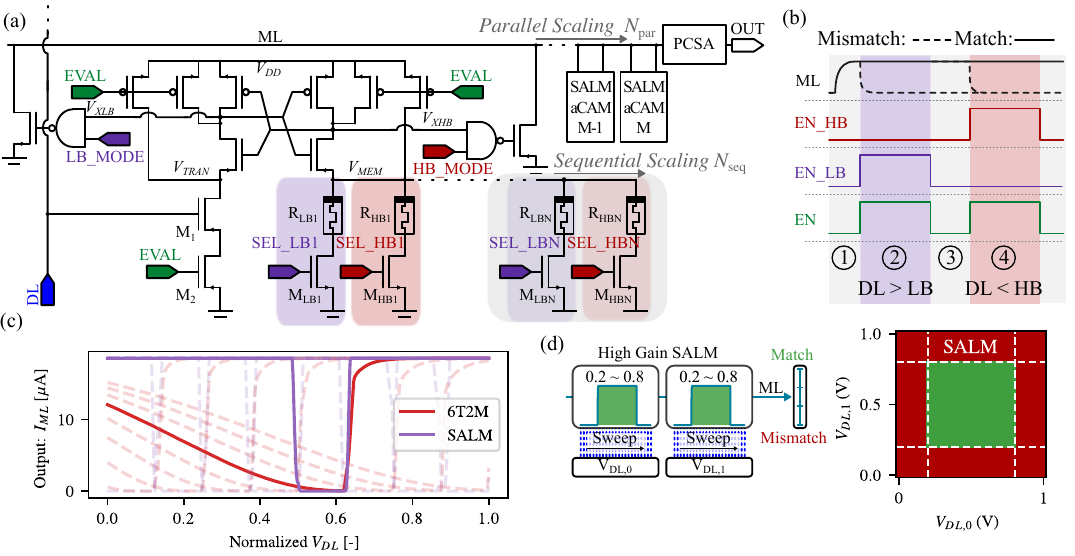}
  \caption{\textbf{(a)} Proposed \ac{SALM} analog CAM architecture featuring $N$ sequential and $M$ parallel memory compartments. \textbf{(b)} Transient control signals for an \ac{SALM} \ac{aCAM} cell with one sequential element. \textbf{(c)} Output \ac{ML} current for eight intervals, illustrating the gain difference (match--mismatch separation) between the proposed \ac{SALM} design and the 6T2M baseline. \textbf{(d)} Experiment showing an input sweep applied to two \ac{SALM} \ac{aCAM} cells with fixed stored boundaries. The resulting output demonstrates the near-ideal match-line (ML) behavior, where the response depends only on the mismatching input and remains well separated near the decision boundary.}
    \label{fig:fig1}
\end{figure*}

\section{Strong Arm Latched Memristor Analog CAM}
\label{sec:SALM}

Our proposed \ac{SALM} \ac{aCAM} design builds on the Strong Arm latch comparator, a well-established topology widely used in \acp{ADC} and memory readout circuits \cite{210039}. The circuit schematic, which consists of a latch and multiple 1T1R structures that store the inequalities, is shown in \figref[a]{fig:fig1}.

In contrast to static divider-based \ac{aCAM} cells such as the 6T2M, the \ac{SALM} evaluates inequalities dynamically through a discharge race between two precharged nodes, $V_{TRAN}$ and $V_{MEM}$, analogous to a dynamic voltage comparator. During evaluation, the regenerative latch amplifies the developing voltage difference and resolves it into one of two stable logic states, thereby providing high gain and intrinsic result latching.

The transient discharge of $V_{TRAN}$ is governed by the analog input voltage applied to the \acf{DL}, whereas $V_{MEM}$ discharges according to the programmed memristor conductance of the currently selected 1T1R pull-down branch. Consequently, the circuit performs a comparison between the applied \ac{DL} voltage and the inequality constraint stored in the memristor, consistent with the principles of \ac{aCAM} operation.

A key architectural feature of the proposed latch-based memristive analog CAM architecture is that a single regenerative latch is reused across multiple memristive discharge paths. Detailed architectural considerations are provided in \secref{sec:SALM_architecture}. The bottom selection switches determine which 1T1R branch connects $V_{\mathrm{MEM}}$ to $GND$ during the evaluation phase. The selected branch corresponds to either a lower-bound or an upper-bound inequality. We denote the corresponding branch-selection signals as \texttt{SEL\_LBN} and \texttt{SEL\_HBN}, respectively. For a given index $n$, exactly one selection signal is asserted at any time, such that only one inequality participates in the comparison.

In combination with the global evaluation clock \texttt{EVAL}, which enables $\textrm{M}_\textrm{2}$, the asserted selection signal activates the corresponding 1T1R branch. As a result, the memristor conductance associated with the selected inequality is connected to $V_{MEM}$ and participates in the comparison operation.

After each evaluation, the latch resolves to either $V_{DD}$ or $V_{SS}$, producing complementary output levels. The comparison polarity, depending on whether a lower-bound or an upper-bound inequality is selected, is determined by a compact gating stage at the top of the latch. This stage maps the complementary latch outputs, denoted as $V_{XLB}$ and $V_{XHB}$, to the \ac{ML} pull-down device. Consequently, the \ac{ML} remains high only if all programmed inequalities are satisfied, whereas any bound violation triggers \ac{ML} discharge. The corresponding control signal transients for a single window evaluation are illustrated in \figref[b]{fig:fig1}.

The circuit was implemented in GlobalFoundries 22\,nm FD-SOI technology and evaluated using SPICE simulations. The resulting performance metrics are summarized in Tab.~\ref{tab:salm_vs_6t2m}. Although the cell area is larger due to the increased transistor count, the overall energy consumption is reduced by eliminating static power during the search operation. Furthermore, while the proposed cell requires two sequential search operations to evaluate \ac{LB} and \ac{HB}, the strong-arm latch enables rapid regeneration, resulting in a shorter evaluation time. In \secref{sec:SALM_architecture}, we demonstrate how the overall \ac{SALM} cell area can be effectively reduced by sharing a single latch across multiple inequalities, resulting in a tradeoff between area efficiency and operational speed.

\begin{table}[t]
\centering
\caption{Key performance comparison between the SALM aCAM and the 6T2M design}
\label{tab:salm_vs_6t2m}
\begin{tabular}{l c c}
\hline
\textbf{Metric} & \textbf{New Design (SALM)} & \textbf{6T2M} \\
\hline
Area (estimated) & 0.498\,$\mu$m$^{2}$ & 0.170\,$\mu$m$^{2}$ \\
Latency & 3\,ns & 3\,ns \\
Energy & 8.8\,fJ & 13.13\,fJ \\
Memory Latch & Yes & No \\
Gain & Large & Small \\
Scalability & Good & Poor \\
\hline
\end{tabular}
\end{table}

\section{Sequential and Parallel Scaling in Latch Shared SALM aCAM Architectures}
\label{sec:SALM_architecture}

The multiplexed architecture comprising a single latch shared across multiple 1T1R elements, as introduced in \secref{sec:SALM}, naturally extends beyond a single window comparison consisting of one \ac{LB} and one \ac{HB} inequality. Additional 1T1R pull-down paths can be integrated and evaluated sequentially, defining a configurable sequential scaling factor $N_{\mathrm{seq}}$ within a single latch instance that enables storage of multiple inequalities per word. Alternatively, the complete cell including the latch can be replicated $N_{\mathrm{par}}$ times to support parallel evaluation, as illustrated in \figref[a]{fig:fig1}. The only structural constraint is that the total number of stored inequalities per word satisfies $N_{\mathrm{word}} \leq N_{\mathrm{seq}} \cdot N_{\mathrm{par}}$.

To make these architectural trade-offs explicit, we introduce an analytical model describing how latency, area, and energy scale with sequential and parallel partitioning of the search word. The model is parameterized by circuit-level characteristics extracted from the SALM implementation in Tab.~\ref{tab:salm_vs_6t2m}. We then use it to explore the design space for the given layout and technology and compare the results with the 6T2M baseline. This analysis clarifies the benefits and limitations of sequential versus parallel scaling and provides a structured basis for selecting architectures under technology, application, and system-level constraints, with a view toward future electronic design automation (EDA) integration.

\paragraph{Latency}
The word-level latency scales linearly with the number of sequential evaluations $N_{\mathrm{seq}}$, since each sequential step incurs a fixed evaluation time associated with precharging and resolving the latch, denoted as $T_{\mathrm{latch}}$.
\begin{equation}
T_{\mathrm{word}} = N_{\mathrm{seq}} \, T_{\mathrm{latch}}.
\label{eq:latency_word}
\end{equation}
The model results are presented in \figref[a]{fig:results_SALM_architecture}

\paragraph{Area}
The area model is not derived from a detailed layout but serves as a first-order estimate to capture scaling trends. The total \ac{SALM} area is dominated by the latch footprint and the number of associated 1T1R elements. Increasing $N_{\mathrm{seq}}$ amortizes the latch area across a larger number of stored inequalities, since the 1T1R structure is significantly smaller than the latch itself. For a parallel replication factor $N_{\mathrm{par}}$, the estimated word-level area is therefore
\begin{equation}
A_{\mathrm{word}} = N_{\mathrm{par}} \left( A_{\mathrm{latch}} + N_{\mathrm{seq}} A_{\mathrm{1T1R}} \right).
\label{eq:area_word}
\end{equation}

A comparison between the conventional 6T2M design and the proposed \ac{SALM} architecture is shown in \figref[b]{fig:results_SALM_architecture}. Although the area of a single \ac{SALM} cell exceeds that of a 6T2M cell, amortizing the latch across multiple sequential inequalities reduces the effective area per word as $N_{\mathrm{seq}}$ increases. Consequently, for sufficiently large $N_{\mathrm{seq}}$, the \ac{SALM} architecture can achieve a smaller word-level area than the 6T2M baseline.

\paragraph{Energy}
The energy model assumes an early termination search mode. During sequential evaluation, further comparisons become unnecessary once a mismatch is detected and the \ac{ML} is discharged, since the corresponding word can no longer match. The expected energy consumption is therefore determined by the probability of encountering a mismatch and by the expected number of sequential evaluation steps that are executed. Both quantities strongly depend on the input data distribution and the target application.

As a simplified statistical model, we assume that each stored inequality check has an independent mismatch probability $P_{\mathrm{mm}}$. In \secref{sec:re_inference}, we evaluate this assumption using representative application-level datasets. In a single sequential step, $N_{\mathrm{par}}$ comparisons are evaluated in parallel. A step is classified as a mismatch if at least one of these parallel comparisons mismatches, yielding the per-step mismatch probability
\begin{equation}
P_{\mathrm{mm,par}} = 1 - \left( 1 - P_{\mathrm{mm}} \right)^{N_{\mathrm{par}}}.
\label{eq:pmm_par}
\end{equation}

Given $P_{\mathrm{mm,par}}$, the sequential evaluation can be modeled as a truncated geometric process. Each step represents an independent Bernoulli trial with mismatch probability $P_{\mathrm{mm,par}}$, and the evaluation terminates either upon the first detected mismatch or after all $N_{\mathrm{seq}}$ stages have been executed. The resulting expected search depth is therefore
\begin{equation}
\mathbb{E}[K] =
\dfrac{1 - \left( 1 - P_{\mathrm{mm,par}} \right)^{N_{\mathrm{seq}}}}{P_{\mathrm{mm,par}}}.
\label{eq:expected_steps}
\end{equation}

Each sequential step consumes the energy required to evaluate $N_{\mathrm{par}}$ cells in parallel, equal to $N_{\mathrm{par}} E_{\mathrm{cell}}$. The expected word-level energy per lookup thus becomes
\begin{equation}
E_{\mathrm{word}} = N_{\mathrm{par}} \, E_{\mathrm{cell}} \, \mathbb{E}[K].
\label{eq:energy_word}
\end{equation}

\begin{figure}[htb!]
    \centering
    \includegraphics[width=\singlefigwidth]{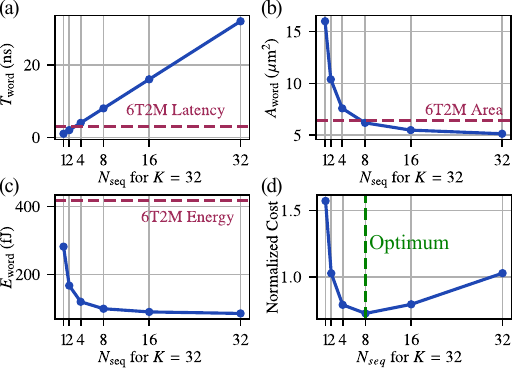}
    \caption{Impact of sequential to parallel architectural partitioning on the performance of a single \ac{aCAM} row with a word size of 32. Shown are (a) latency, (b) estimated chip area per row/word, (c) search energy, and (d) a combined cost function integrating normalized latency, area, and energy metrics.}
    \label{fig:results_SALM_architecture}
\end{figure}

\figref[c]{fig:results_SALM_architecture} shows the expected energy consumption predicted by this model as a function of the number of sequential elements.


Latency, area, and energy form a coupled design space in which improvements in one metric typically come at the expense of another. The optimal operating point therefore depends on the relative importance assigned to these metrics. It is further influenced by technology and layout choices, which directly affect $A_{\mathrm{latch}}$ and $T_{\mathrm{step}}$, as well as by the target workload, since the mismatch statistics determine the expected energy consumption in \eqref{eq:energy_word}.

To enable a fair comparison between different design points, we normalize all metrics to obtain dimensionless quantities and combine them into a unified cost function, as defined in \eqref{eq:cost}.

A single scalar objective is obtained as the weighted sum
\begin{equation}
C = w_E E_{\mathrm{norm}} + w_L T_{\mathrm{norm}} + w_A A_{\mathrm{norm}},
\label{eq:cost}
\end{equation}
with $w_E + w_L + w_A = 1$.

The weights reflect the application-driven priorities. Minimizing $C$ selects an architecture that best matches these preferences, while the individual normalized metrics preserve full transparency of the latency, energy, and area trade-off.

\figref[d]{fig:results_SALM_architecture} shows an optimization example using the parameter set \(w_E, w_L, w_A = [0.45, 0.05, 0.5]\). The resulting convex curve indicates that the optimal solution is obtained for \(n_{\mathrm{seq}} = 8\).

\section{Circuit-Accurate Behavioral Model}
\label{sec:beh_model}

To enable large-scale inference simulations, we developed a circuit-accurate \ac{aCAM} behavioral model that captures the essential dynamics responsible for \ac{ML} crosstalk, as full SPICE simulations are prohibitively slow for system-level studies. The model is entirely constructed from \acp{LUT} extracted through transistor-level SPICE simulations using the GlobalFoundries 22~FD-SOI technology, ensuring an accurate representation of the underlying circuit behavior. The framework is general and can be applied to any \ac{aCAM} architecture. The complete implementation and documentation are publicly available on GitHub.

\begin{figure*}[htb!]
    \centering
    \includegraphics[width=\textwidth]{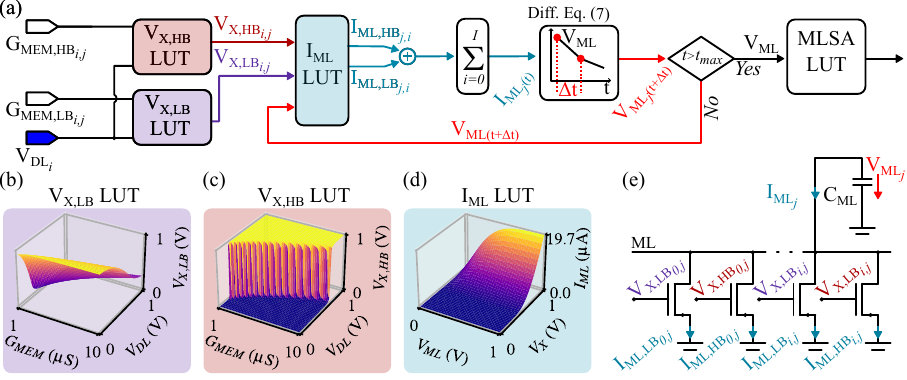}
    \caption{
    (a) Flow diagram of the behavioral model, including the use of lookup tables (\acp{LUT}) and the iterative loop used to solve the differential equation governing the discharge of the match-line (\ac{ML}) capacitor.
    (b,c) \acp{LUT} describing the inequality-check output voltages $V_X$ for the lower-bound (LB) and upper-bound (HB) comparisons as a function of the memristor conductance $G_{MEM}$ and the search voltage $V_{DL}$.
    (d) \ac{LUT} mapping the match-line current $I_{ML}$ as a function of the match-line voltage $V_{ML}$ and the comparator output voltage $V_X$.
    (e) NOR-type match-line configuration with multiple parallel pull-down transistors driven by the voltages $V_X$, illustrating the aggregation of cell contributions on the \ac{ML}.
    }
   \label{fig:beh_model}
\end{figure*}

As illustrated in \figref[a]{fig:beh_model}, the behavioral model consists of three distinct \acp{LUT}. The first two tables provide the comparator output node voltages $V_{X,LB}$ and $V_{X,HB}$ as functions of the programmed memristor conductance and the applied data-line voltage $V_{\mathrm{DL}}$. These voltage-transfer characteristics primarily capture the effective voltage gain of the underlying \ac{aCAM} circuitry and therefore determine the separation between match and mismatch drive levels at the node $V_X$. The third \ac{LUT} stores the current--voltage characteristic of the match-line pull-down device, returning the cell discharge current $I_{\mathrm{ML}}$ as a function of the instantaneous match-line voltage $V_{\mathrm{ML}}$ and the corresponding drive voltage $V_X$. This formulation is tailored to the widely used NOR-type match line, where multiple pull-down devices are connected in parallel. The total match-line current is obtained by summing the individual cell currents according to Kirchhoff's current law. Since the pull-down device operates in the nonlinear regime, $I_{\mathrm{ML}}$ depends on $V_{\mathrm{ML}}$ and evolves during discharge. Consequently, the model updates $V_{\mathrm{ML}}$ iteratively in discrete time steps $\Delta t$ according to
\begin{equation}
V_{\mathrm{ML}}(t_i + \Delta t) = V_{\mathrm{ML}}(t_i) - \frac{I_{\mathrm{ML}}(t_i)}{C_{\mathrm{ML}}} \, \Delta t ,
\end{equation}

until the maximum evaluation time $t_{\mathrm{max}}$ is reached. This iterative approach reproduces the dynamic discharge behavior of the match line with high fidelity while remaining several orders of magnitude faster than full SPICE simulation.

Together, these tables enable the model to accurately reproduce the transient match-line behavior without requiring a full SPICE simulation. \figref[a\&b]{fig:beh_model_vs_spice} shows a comparison between the SPICE-simulated output characteristics of an \ac{aCAM} cell and the corresponding behavioral model across the complete input parameter sweep of $G_{MEM}$ and $V_{DL}$, using a conductance window of \valunit{1}{\mu S}.

\begin{figure}[htb!]
    \centering
    \includegraphics[width=\singlefigwidth]{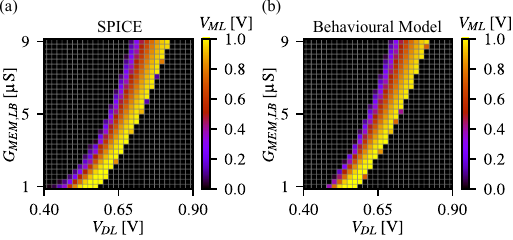}
    \caption{Comparison of match-line discharge between (a) transistor-level SPICE and (b) the circuit-accurate behavioral model across the full sweep of search voltage $V_{\mathrm{DL}}$ and programmed lower-bound conductance $G_{\mathrm{MEM,LB}}$. The colormap shows the match-line voltage $V_{\mathrm{ML}}$ at the end of the evaluation window for a conductance window of \valunit{1}{\mu S}, demonstrating close agreement of the behavioral model with SPICE over the complete operating range.}
    \label{fig:beh_model_vs_spice}
\end{figure}


\section{Random Forest Inference Using aCAM Arrays}
\label{sec:re_inference}

We employ the \texttt{sklearn} framework to train a conventional Random Forest classifier, which serves as the algorithmic baseline. After training, the X--TIME compiler~\cite{10753423} extracts from the fitted model the set of analog comparison intervals associated with every decision node. These intervals correspond directly to the low-- and high--bound inequalities that can be programmed into an \ac{aCAM} array. The compiler therefore produces, for each tree branch, the complete set of \ac{aCAM} parameters that encode the decision structure of the Random Forest.

The generated parameters are mapped onto our behavioral \ac{aCAM} model, which internally uses technology-specific \acp{LUT} to represent the input--output characteristics of each considered \ac{aCAM} cell architecture. These tables capture the dependence of the comparator outputs and pull-down currents on both the applied data-line voltage and the programmed memristor conductances, ensuring that array-level effects such as \ac{ML} crosstalk are accurately reproduced during inference.

The dimensionality of each dataset, listed in Tab.~\ref{tab:dataset}, determines the length of the search vector applied to aCAM rows. In practice, this feature dimension often exceeds the maximum feasible number of \ac{aCAM} cells that can be connected to a single \ac{ML} without compromising accuracy or read latency. To remain within realistic hardware constraints, we therefore tile the computation by limiting each \ac{ML} segment to 64 cells. A search vector exceeding this length is partitioned into multiple tiles, each evaluated independently by its own \ac{ML} and \ac{MLSA}. Every tile produces a binary partial-match result that indicates whether the corresponding feature subset satisfies all programmed inequalities.

Finally, a digital aggregation stage collects the partial-match results of all tiles belonging to the same decision path. Only if all tiles return a match does the full feature vector satisfy the original inequality encoded by the decision-tree node. This hierarchical evaluation replicates exactly the semantics of the software Random Forest while respecting the architectural constraints and analog behavior of the underlying \ac{aCAM} hardware.


\figref[a]{fig:re_inference} summarizes the inference results, comparing the achieved accuracy across multiple datasets for three distinct \ac{aCAM} circuit architectures. For datasets with low feature dimensionality, all architectures achieve accuracy values close to the ideal baseline, as the corresponding \ac{ML} lengths remain short and crosstalk effects are minimal. As the feature dimension grows, however, the accuracy gap between the architectures increases. In these larger models, the \ac{SALM} design consistently provides the highest accuracy, reflecting its superior gain and reduced susceptibility to \ac{ML} crosstalk. 

\figref[b]{fig:re_inference} illustrates the dependence of inference accuracy on the \ac{ML} segment length, which represents the number of \ac{aCAM} cells connected to a single \ac{ML} in parallel. Using the Gas Concentration dataset as an example, we observe that increasing the array tile size leads to a monotonic accuracy degradation, with different rates for each architecture. The \ac{SALM} design again demonstrates the highest robustness, degrading significantly more slowly than the \ac{6T2M} design.

\begin{figure}[htb!]
    \centering
    \includegraphics[width=\singlefigwidth]{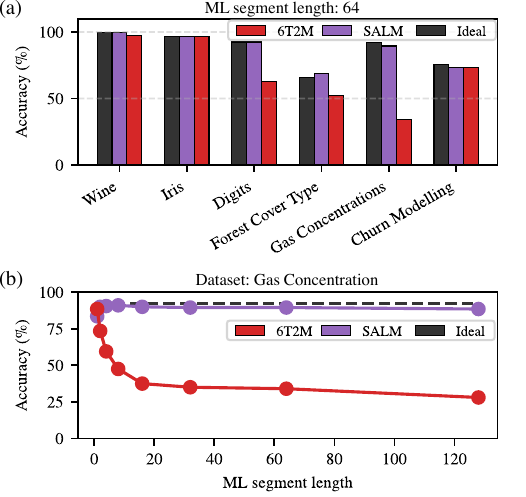}
    \caption{(a) Comparison of inference accuracy between an ideal \texttt{scikit-learn} predictor and the \ac{aCAM} hardware implementation for different models. (b) Inference accuracy versus \ac{ML} segment length for the gas concentration dataset}
    \label{fig:re_inference}
\end{figure}

\paragraph{Dataset Dependent Energy Savings in Sequential Architecture Scaling}

As described in \secref{sec:SALM_architecture}, where we modeled the energy consumption through a probability distribution of mismatch probability per feature (see Eqs.~\ref{eq:pmm_par}--\ref{eq:energy_word}), we now evaluate the real-world behavior during inference on representative application-level datasets.

We compute mismatch statistics directly from the compiled tree models and measure the effective sequential search depth observed during inference, defined as the number of features evaluated before the first mismatch aborts a search for a given key. In this experiment, we investigate how many sequential elements are required to achieve a specified level of energy reduction.

The results indicate that already three sequential elements per search are sufficient to reduce the energy consumption by approximately 50\% for most datasets. This corresponds to a threefold increase in latency, while simultaneously reducing energy by half and significantly decreasing the required latch replication and area. To achieve an energy reduction of approximately 80\%, approximately six sequential evaluations are required for most workloads.

\begin{table}[h]
\centering
\caption{Dataset Characteristics and Sequential Scaling Thresholds Required to Achieve 50\% and 80\% Energy Reduction}
\label{tab:dataset}
\begin{tabular}{lcccc}
\hline
\textbf{Dataset} & \textbf{Features} & \textbf{Classes} & $\mathbf{N_{\mathrm{seq},50\%}}$ & $\mathbf{N_{\mathrm{seq},80\%}}$ \\
\hline
Churn~\cite{churn_modelling_kaggle} & 965 & 2 & 3 & 6 \\
Digits~\cite{alpaydin1998optical} & 60 & 10 & 3 & 6 \\
Forest Cover~\cite{blackard1999cover} & 47 & 7 & 3 & 6 \\
Gas~\cite{vergara2012gas} & 129 & 6 & 3 & 6 \\
Iris~\cite{fisher1936iris} & 4 & 3 & 4 & -- \\
Wine~\cite{forina1991wine} & 13 & 3 & 3 & -- \\
\hline
\end{tabular}
\end{table}

\section{Conclusion}

This work addresses the fundamental energy and scalability limitations of the 6T2M \ac{aCAM} design. We identified static search power, limited voltage gain, and cumulative match-line crosstalk in 6T2M cells as key bottlenecks that restrict analog precision and large-scale deployment.

To overcome these limitations, we introduced the Strong Arm Latched Memristor (SALM) aCAM cell, which replaces static voltage division with a dynamic current-race comparator. The latch-based comparison mechanism eliminates static search current, provides high regenerative gain, and intrinsically stores the comparison result. Circuit-level evaluation in 22\,nm FD-SOI demonstrates a reduction of approximately 33\% in read energy at identical latency, while significantly improving match--mismatch separation and robustness against match-line crosstalk.

Beyond the cell level, we exposed architectural degrees of freedom by enabling sequential reuse and parallel replication of latch-shared compartments. We derived analytical latency, area, and energy models that enable workload-aware design-space exploration. Dataset-level evaluation shows that already three sequential evaluations are sufficient to reduce search energy by approximately 50\% for most workloads, while maintaining inference accuracy close to the software baseline. This highlights the importance of workload-dependent architecture optimization rather than fixed parallel array sizing.

To support large-scale system studies, we further developed a circuit-accurate behavioral model derived from SPICE lookup tables, enabling efficient inference simulations while preserving match-line discharge dynamics. Integration into the X-TIME compiler framework allowed quantitative comparison across multiple aCAM architectures and demonstrated that SALM maintains near-software accuracy even at large match-line lengths where baseline designs degrade rapidly.

Overall, latch-based analog CAMs provide a scalable and energy-efficient foundation for associative memory accelerators. Future work will focus on large-scale array integration, adaptive workload-aware mapping strategies, and the exploration of latch-based aCAM primitives in emerging in-memory AI accelerators.

\section*{Acknowledgment}

This work was supported by Hewlett Packard Enterprise and Forschungszentrum Jülich.

\bibliographystyle{IEEEtran}
\bibliography{literature}

\end{document}